\documentclass{iaus}
\usepackage{graphics}

  \checkfont{eurm10}
  \iffontfound
    \IfFileExists{upmath.sty}
      {\typeout{^^JFound AMS Euler Roman fonts on the system,
                   using the 'upmath' package.^^J}%
       \usepackage{upmath}}
      {\typeout{^^JFound AMS Euler Roman fonts on the system, but you
                   dont seem to have the}%
       \typeout{'upmath' package installed. iaus.cls can take advantage
                 of these fonts,^^Jif you use 'upmath' package.^^J}%
      }
  \else
  \fi

  \checkfont{msam10}
  \iffontfound
    \IfFileExists{amssymb.sty}
      {\typeout{^^JFound AMS Symbol fonts on the system, using the
                'amssymb' package.^^J}%
       \usepackage{amssymb}%

      }{}
  \fi

  \IfFileExists{amsbsy.sty}
    {\typeout{^^JFound the 'amsbsy' package on the system, using it.^^J}%
     \usepackage{amsbsy}}
    {}

\newsavebox{\astrutbox}
\sbox{\astrutbox}{\rule[-5pt]{0pt}{20pt}}

\title[Outskirts of Galaxy Clusters: intense life in the suburbs]
      {Breaking the mass / anisotropy degeneracy 
	in the Coma cluster}
\author[G. A. Mamon {\it et al.\/}]%
{Gary A. Mamon$^1$,
Ewa L. {\L}okas$^2$
\and Teresa Sanchis$^3$}

\affiliation{$^1$IAP, Paris, FRANCE email: {\tt gam@iap.fr}\\[\affilskip]
$^2$Copernicus Center, Warsaw, POLAND email: {\tt
lokas@camk.edu.pl}\\[\affilskip]
$^3$Dep. di Astronomia i Meteorologia, Univ. de Barcelona, SPAIN email: {\tt
tsanchis@am.ub.es}}

\pubyear{2004}
\volume{195}
\pagerange{1--5}
\date{?? and in revised form ??}
\jname{Outskirts of Galaxy Clusters: intense life in the suburbs}
\editors{A. Diaferio, ed.}

\begin{document}

\maketitle

\begin{abstract}
We provide the first
direct lifting of the mass/anisotropy degeneracy for a cluster of galaxies,
by jointly 
fitting the line of sight velocity dispersion and kurtosis profiles of the
Coma cluster,
assuming an NFW tracer density profile, a generalized-NFW dark matter profile
and a constant anisotropy profile.
We find that the orbits in Coma 
must be quasi-isotropic, and find a mass consistent
with previous analyses, but a concentration parameter 50\% higher than 
expected in cosmological $N$-body simulations.
We then test the accuracy of our method on realistic non-spherical systems
with substructure and streaming motions, by applying it to the ten most
massive structures in a  cosmological $N$-body simulation.
We find that our method yields fairly accurate results on average (within
20\%), although with a wide variation (factor 1.7 at 1$\,\sigma$) 
for the concentration parameter, with decreased accuracy and efficiency when
the projected mean velocity is not constant with radius.
\end{abstract}

\firstsection 
\section{Introduction}
The kinematical (``Jeans'') 
analyses of near-spherical structures using the spherical
Jeans equation for stationary systems:
\[
    \frac{\rm d}{{\rm d} r}  (\nu \sigma_r^2) + \frac{2 \beta}{r} \nu
	\sigma_r^2 = - \nu \frac{G\,M}{r^2} \ ,
\]
where $\nu$ is the 3D density distribution of the tracer population, 
suffer from the fact that this one
equation involves two unknowns: the mass distribution $M(r)$ and the velocity
anisotropy $\beta = 1 - \sigma_\theta^2/\sigma_r^2$ (we make use of the
symmetry of spherical systems yielding $\sigma_\phi = \sigma_\theta$).
In some cases of simple anisotropy profiles $\beta(r)$, the Jeans equation
can be inverted to yield the radial dispersion as a single integral over
radii of $\nu M$ times some kernel, and one can insert this solution into the
equation that links the line-of-sight velocity dispersion to the radial one,
to obtain after some algebra, the line-of-sight dispersion as a single
integral over $\nu M$ and some other kernel.

In what follows, we attempt to lift the mass/anisotropy degeneracy by adding
a second equation, namely the 4th order Jeans equation for constant
anisotropy ({\L}okas 2002)
\[
        \frac{\rm d}{{\rm d} r}  (\nu \overline{v_r^4}) + \frac{2 \beta}{r} \nu
	\overline{v_r^4} + 3 \nu \sigma_r^2 \frac{{\rm d} \Phi}{{\rm d} r} =0
	\ ,
\]
where we express the line of sight \emph{kurtosis} as a double integral over
$\nu M$ and $M$ and a third kernel (the details are given in {\L}okas \&
Mamon 2003).

\section{Kinematic analysis of the Coma cluster}
We apply our formalism to the nearby Coma cluster, for which we've extracted
from {\sf NED} the positions and line-of-sight velocities of 967 galaxies,
among which 355 early-type galaxies whose kinematics we analyze here (we omit
spiral galaxies, which are expected to have a non negligible mean radial
velocity caused by their infall and subsequent transformation at cluster
pericenter). Our mass model includes 1) a stellar component, which we find is
well fit by an NFW (Navarro et al. 1996) model; 
2) a dark matter component, which we assume to
be a generalized NFW model with arbitrary inner slope:
\begin{equation}
\rho(r) \propto (r/a)^{-\alpha}\,(1+r/a)^{\alpha-3} \ ,
\label{rhoofr}
\end{equation}
where 
$a$ is the scale radius of the dark matter (in general different from that of
the luminous distribution), and $\alpha=1$ for the NFW model. The profiles
differ by their inner slope $-\alpha$
but have a common outer limiting behavior of $r^{-3}$;
and 3) an isothermal
gas component, taken from the {\sf ROSAT} X-ray
observations by Briel et al. (1992).

\begin{figure}
\resizebox{0.98\hsize}{!}{\includegraphics{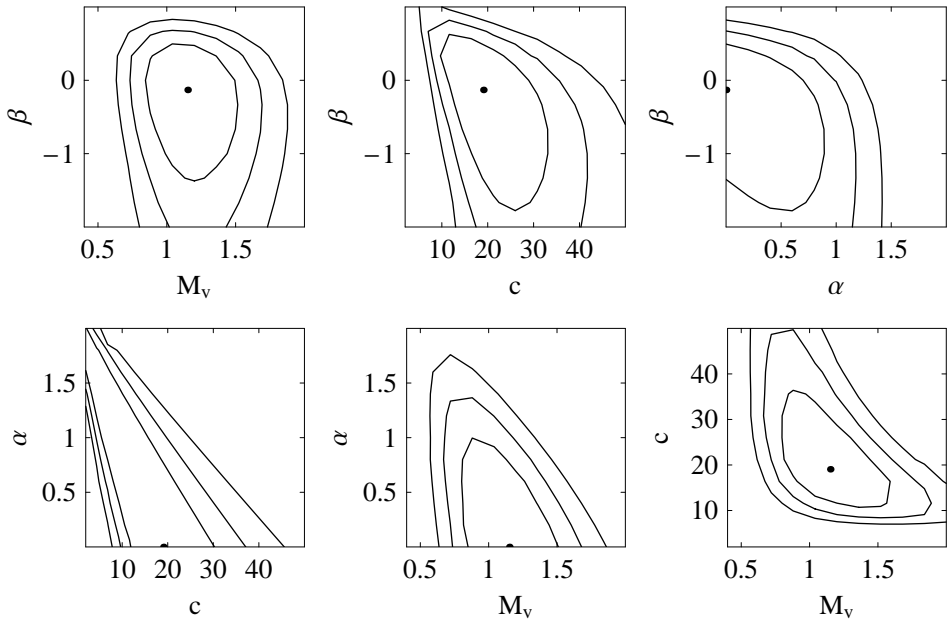}}
\caption{{\small Probability contours (1, 2, and $3\,\sigma$)
for joint fits of $\sigma_{\rm los}$
and $(\log \kappa_{\rm los})^{1/10}$ (the two are found to be independent).
\emph{Circles} indicate the best-fitting parameters.
The mass is in units of $10^{15} M_\odot$.}}
\label{allplots}
\end{figure}

Using 7 radial bins at $R<80'$, we jointly
fit our four parameters, mass $M_v$ within the virial radius $r_v$, 
velocity anisotropy $\beta$, inner slope $\alpha$ of the dark matter density
profile, and dark matter concentration $c=r_v/a$. Contrary to the case
when only velocity dispersion is studied, the minimization procedure now
converges (except for a concentration / inner slope degeneracy). 
The best fits are shown in Figure~\ref{allplots}.
The best fit parameters are
$M_v = 1.2 \times 10^{15} M_\odot$ (corresponding to dark matter
virial radius $r_v=92' = 2.7$ Mpc), $\beta=-0.13$, $\alpha=0$ and $c=19$
with $\chi^2/N = 6.1/10$. 
In other words, our best fit models are isotropic and the generalized dark
matter model of equation~(\ref{rhoofr}) provides excellent fits to the data.

\begin{figure}
\resizebox{0.9\hsize}{!}{\includegraphics{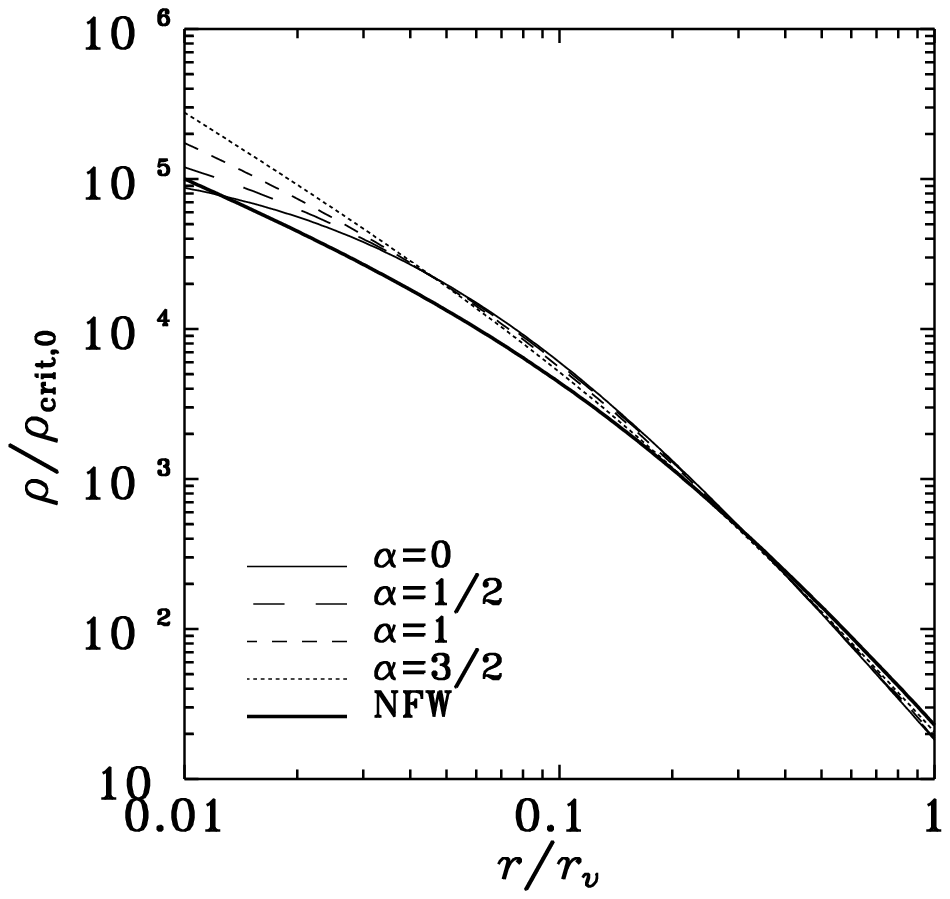}\includegraphics{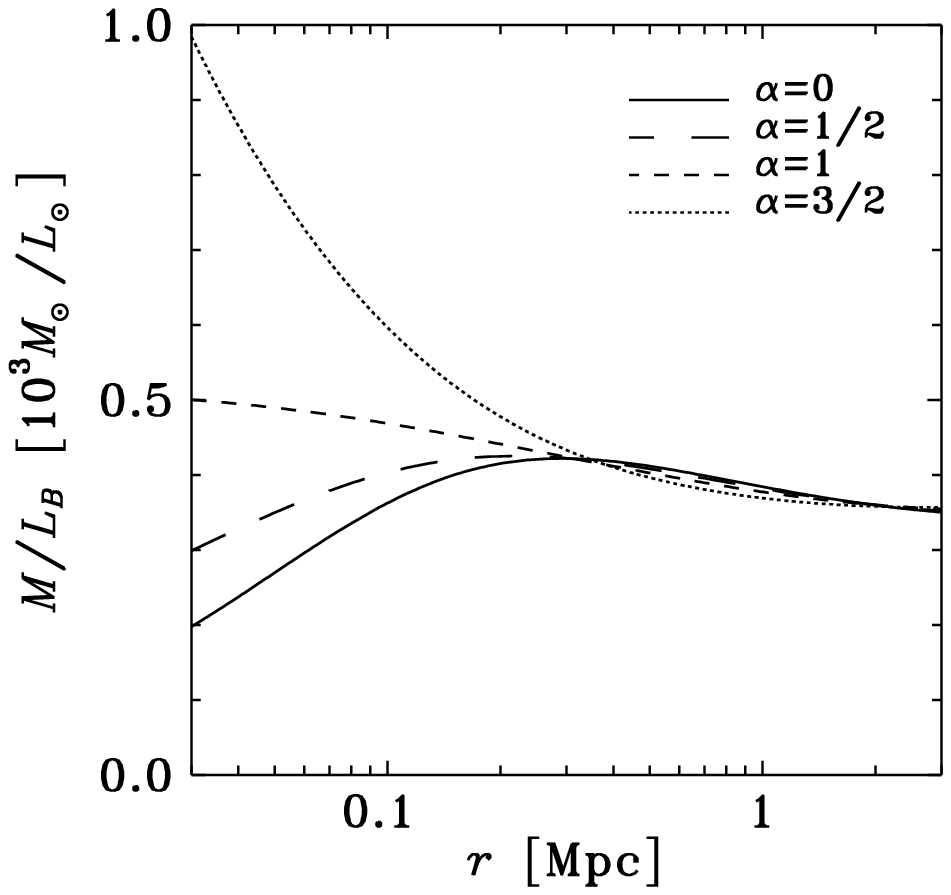}}
\caption{{\small Best fitting dark matter profiles with different inner slopes
  (\emph{left  panel}) and their associated mass-to-light ratios
  (\emph{right panel}), for  
  different $\alpha$ increasing upwards from 0 to 1.5 (with $c=19$, 14, 9 and
5, respectively).
The dark matter virial radii are $r_v=2.7$ Mpc for all models shown.
The \emph{heavy solid curve} in the left plot 
indicates the best NFW model ($c=6$) from cosmological
$N$-body simulations, which is still (marginally) consistent with the data.
}}
\label{profs}
\end{figure}

Our best fit radial profiles 
for each half-integer value of $\alpha$ are shown in
Figure~\ref{profs}.
The best fit density
profiles are indistinguishable beyond $0.03\,r_v$, and the
outer mass profiles are very close to the infall 
estimates of Geller, Diaferio \& Kurtz (1999).
The density profiles
 are 50\% more concentrated than in
structures of comparable mass within cosmological $N$-body simulations in a
$\Lambda$CDM Universe (Bullock et al. 2001). Interestingly, the 
$M/L$ profiles have 
nearly constant slope for the NFW case ($\alpha=1$).

Our analysis differs from previous kinematical analyses (e.g. 
Merritt 1987; den Hartog \& Katgert 1996; Carlberg et al. 1997;
van der Marel et al. 2000; Biviano \& Girardi 2003; Katgert et al. 2004),
as 1) we have, for a single cluster, a larger sample of galaxies, which,
given their early morphological type, should be in dynamical equilibrium in
the cluster potential;
2) we remove pairs from the
computation of the velocity moments;
3) we include kurtosis in the analysis;
4) we model the
dark matter distribution using a generalized formula inspired by the
results of cosmological $N$-body simulations;
5) we include hot gas.

In comparison to studies based upon stacking of many clusters, our analysis
of the Coma cluster benefits from not having to introduce errors in any
stacking procedure, and from a cleaner removal of interlopers.
On the other hand,
the analyses of stacked clusters have the advantage of averaging out
particular inhomogeneities of individual clusters such as Coma, expected in
hierarchical scenarios of structure formation and observed in Coma,
both in
projected space (Fitchett \& Webster 1987; Mellier et al. 1988; Briel et
al. 1992) and velocity space (Colless \& Dunn 1996; Biviano et al. 1996).

\section{Kinematic analysis of simulated clusters}

In our Jeans analysis of the  Coma cluster, we assumed that the cluster was
spherical, with no substructure nor streaming motions, and that projection
effects were not serious.
We test (details in Sanchis, {\L}okas \& Mamon 2004)
these assumptions using an $\Lambda$CDM cosmological $N$-body (tree-code) 
simulation, run by
Ninin (1999), with $256^3$  particles, and
with $\Omega_m=1/3$, $\Omega_\lambda=2/3$, $h=2/3$, and $\sigma_8=0.88$
(details in
Hatton et al. 2003).

We analyze each of the 10 most massive halos along 3 axes: the major axis, 
an orthogonal one
and an intermediate one.
\begin{figure}
\resizebox{0.84\hsize}{!}{\includegraphics{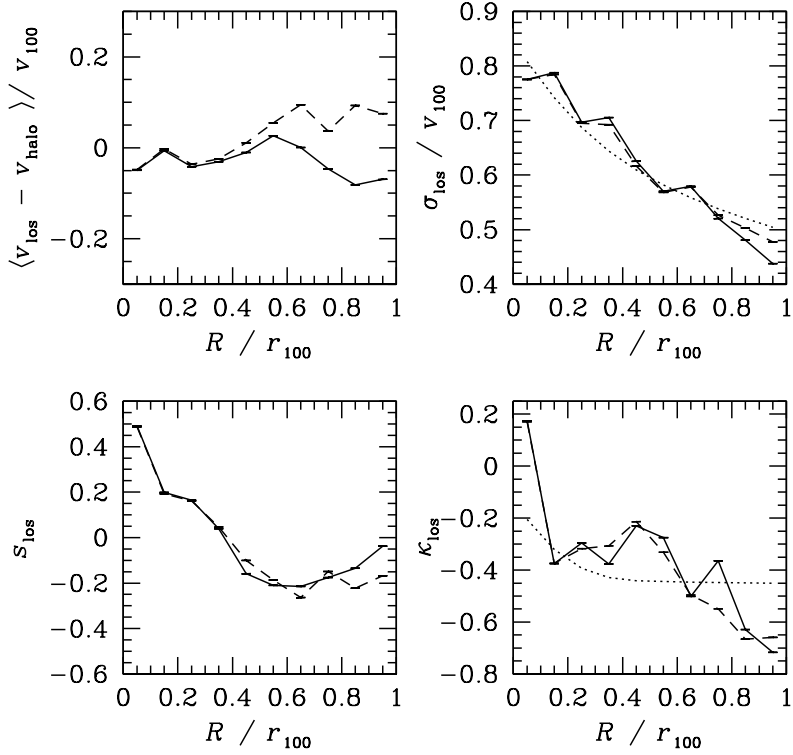}}
\caption{
Projected velocity moments of the dark matter particles in halo 1 measured
along the principal axis. The \emph{upper left panel} shows the mean
line-of-sight velocity with respect to the velocity of the center of the halo
in units of the circular velocity at $r_{\rm 100}$. The \emph{upper right
panel} gives the line-of-sight velocity dispersion in the same units. The two
\emph{lower panels} give the skewness (\emph{left}) and kurtosis
(\emph{right}). In each panel the \emph{solid line} shows results for
particles lying inside the sphere of radius $r_{\rm 100}$, while the
\emph{dashed line} is for all particles. The \emph{dotted curve} shows the
fits obtained from the Jeans equations.}
\label{mom0}
\end{figure}
\begin{figure}
\resizebox{\hsize}{!}{\includegraphics{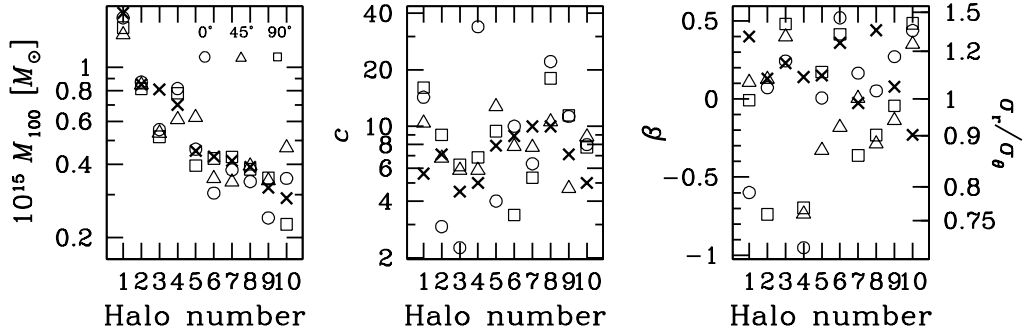}}
\caption{{\small 
Fitted (on projected data) values (using 40 particles per bin) 
of virial mass $M_{\rm 100}$,
concentration parameter $c$ and anisotropy $\beta$ of the ten halos for the
three directions of observation with respect to the major axis of each halo:
$0^\circ$ (\emph{circles}), $45^\circ$ (\emph{triangles})
and $90^\circ$ (\emph{squares}). The
values measured on 3D data are shown with \emph{crosses}.}}
\label{fits}
\end{figure}
Figure~\ref{mom0} shows the radial profiles of 
line-of-sight velocity moments (mean,
dispersion, skewness and kurtosis) along the principal axis.
These profiles
display statistically significant
radial variations caused by substructures with radial streaming motions.
Moreover, the moments vary considerably from halo to halo and for the 3
different projections of a given halo. 
Note also that the projected mean velocity differs substantially between all
the particles in the projected cylinder and the subset within the virial
sphere.

Figure~\ref{fits} shows the results of our fits from our Jeans analysis  
(\emph{open symbols}) in
comparison with our fits to the 3D data (\emph{crosses}) and
Table~\ref{stats} summarizes the statistical accuracy of our fitting procedure.
\begin{table}
\tabcolsep 2pt
\begin{center}
\begin{tabular}{cccccccccccc}
\hline
\hline
Particles & \multicolumn{2}{c}{$\Delta \log M_{100}$} &
& \multicolumn{2}{c}{$\Delta\log c$} &
& \multicolumn{2}{c}{$\Delta\beta$} &
& \multicolumn{2}{c}{$\Delta \log (\sigma_{r}/\sigma_{\theta})$} \\[3pt]
\cline{2-3}
\cline{5-6}
\cline{8-9}
\cline{11-12}
per bin & mean & $\sigma$ &
& mean & $\sigma$ &
& mean & $\sigma$ &
& mean & $\sigma$ \\
\hline
All & --0.03 &  0.09 && 0.20 & 0.18 && --0.78 & 1.04 &&--0.12 & 0.12 \\
40  & --0.07 &  0.10 && 0.08 & 0.24 && --0.20 & 0.48 &&--0.04 & 0.11  \\
\hline
\end{tabular}
\caption{Results of the fitting procedure}
\label{stats}
\end{center}
\end{table}
The virial mass is typically underestimated by $15\pm26\%$, while the 
concentration 
parameter is overestimated by $20\pm74\%$ and the ratio
$\sigma_r/\sigma_\theta$ 
underestimated by $9\%\pm29\%$. Hence, our kinematical analysis yields good
results \emph{on average}, but with a fairly large dispersion in accuracies
(with the most inaccurate estimates for halos with non zero mean line of
sight velocities at some radii).
Halo 4, which departs the most from the 3D measures of concentration and
anisotropy, happens to have important variations in its projected mean
velocity profile, whereas nearly all other halos have normal mean velocity
profiles. This suggests that the joint 
dispersion/kurtosis Jeans analysis is adequate for structures with fairly
constant projected mean velocity profiles.
Given the large dispersion on the concentration parameter, our high
value of $c$ for Coma is now consistent with the 1.5 times lower value
expected from cosmological $N$-body simulations.


\end{document}